\documentclass[twocolumn,superscriptaddress,floatfix,preprintnumbers,prl,nofootinbib,hyperref]{revtex4-2}
\pdfoutput=1
\usepackage[colorlinks=true,breaklinks=true]{hyperref}
\usepackage[normalem]{ulem}
\usepackage[utf8]{inputenc}
\hypersetup{allcolors=[rgb]{0.0 0.0 0.6},linkcolor=[rgb]{0.75 0.05 0.05}}
\usepackage{amsmath,amssymb}
\usepackage{bm}
\usepackage{mathtools}
\usepackage[dvipsnames]{xcolor}
\usepackage{float}
\usepackage{letltxmacro}
\LetLtxMacro{\oldcite}{\cite}
\renewcommand{\cite}[1]{\mbox{\oldcite{#1}}}
\long\def\exclude#1{}

\newcommand{\gagg}{g_{a\gamma \gamma}}

\newcommand{\beq}{\begin{equation}}
\newcommand{\eeq}{\end{equation}}

\def\ga{\,\,\raise0.14em\hbox{$>$}\kern-0.76em\lower0.28em\hbox
{$\sim$}\,\,}

\newcommand{\ie}{\emph{i.e.~}}
\newcommand{\eg}{\emph{e.g.~}}

\long\def\exclude#1{}

\newcommand{\GJ}{\rho_{{}_{\rm GJ}}}

\allowdisplaybreaks

\bibpunct{[}{]}{,}{n}{}{,}

\hypersetup{
    colorlinks=true,
    citecolor=[rgb]{.1, .7, .6},
    linkcolor=[rgb]{.2, .55, .95},
    filecolor=magenta,
    urlcolor=[rgb]{.1, .7, .6},
}

\begin{document}

\title{Pulsar Nulling and Vacuum  Radio Emission from Axion Clouds }

\author{Andrea Caputo} \email[]{andrea.caputo@cern.ch}
\affiliation{Department of Theoretical Physics, CERN, Esplanade des Particules 1, P.O. Box 1211, Geneva 23, Switzerland}

\author{Samuel J. Witte}
\email[]{samuel.witte@physics.ox.ac.uk}
\affiliation{Rudolf Peierls Centre for Theoretical Physics, University of Oxford, Parks Road, Oxford OX1 3PU, UK}
\affiliation{Departament de F\'{i}sica Qu\`{a}ntica i Astrof\'{i}sica and Institut de Ciencies del Cosmos (ICCUB) ,
Universitat de Barcelona, Diagonal 647, E-08028 Barcelona, Spain}

\author{Alexander A. Philippov} 
\affiliation{Department of Physics, University of Maryland, College Park, MD 20742, USA}

\author{Ted Jacobson} 
\affiliation{Department of Physics, University of Maryland, College Park, MD 20742, USA}


\begin{abstract}

Non-relativistic axions can be efficiently produced in in the polar caps of pulsars, resulting in the formation of a dense cloud of gravitationally bound axions. Here, we  investigate the interplay between such an axion cloud and the electrodynamics in the pulsar magnetosphere, focusing specifically on the dynamics in the polar caps, where the impact of the axion cloud is expected to be most pronounced. For sufficiently light axions $m_a \lesssim 10^{-7}$ eV, we show that the axion cloud can occasionally screen the local electric field responsible for particle acceleration and pair production, inducing a periodic nulling of the pulsar's intrinsic radio emission. At larger axion masses, the small-scale fluctuations in the axion field tend to suppress the back-reaction of the axion on the electrodynamics; however, we point out that the incoherent oscillations of the axion in short-lived regions of vacuum near the neutron star surface can  produce a narrow radio line, which provides a complementary source of radio emission to the plasma-resonant emission processes identified in previous work. While this work focuses on the leading order correction to pair production in the magnetosphere, we speculate that there can exist dramatic deviations in the electrodynamics of these systems when the axion back-reaction becomes non-linear. 
\end{abstract}

\maketitle

The QCD axion and axion-like particles are amongst the most compelling candidates for physics beyond the Standard Model, owing to their ability to resolve major outstanding problems in modern physics (such as the question of why QCD conserves charge-parity symmetry~\cite{Peccei:1977hh, Peccei:1977ur, WeinbergAxion, WilczekAxion}, and the nature of dark matter~\cite{Preskill:1982cy, Fischler1982}), and the fact that they arise ubiquitously in  well-motivated high-energy theories such as String Theory~\cite{Arvanitaki:2009fg, Witten:1984dg, Cicoli:2012sz, Conlon:2006tq, Svrcek:2006yi}. 

Significant progress has been made in understanding how to search for and detect axions, with a majority of experiments and proposals attempting to observe the coupling of axions to electromagnetism (see \eg~\cite{Irastorza:2018dyq} for a review). One of the particularly promising ideas that has been put forth to indirectly detect axions is to point radio telescopes at neutron stars -- the idea being that the large magnetic fields and ambient plasma filling the magnetospheres can dramatically enhance axion-photon interactions. Various  observational signatures arising in these systems have been identified~\cite{Pshirkov:2007st,Hook:2018iia, Huang:2018lxq, Leroy:2019ghm, Safdi:2018oeu, Battye:2019aco, Foster:2020pgt, Foster:2022fxn, Witte2021,millar2021axionphotonUPDATED,battye2021robust,Witte:2022cjj,Battye:2023oac,Tjemsland:2023vvc, Prabhu:2021zve,MainGapPaper, BoundStates, Iwazaki:2014wka,Bai:2017feq,Dietrich:2018jov,Prabhu:2020yif,Edwards:2020afl,Buckley:2020fmh,Nurmi:2021xds,Bai:2021nrs,Witte:2022cjj,Prabhu:2023cgb,BoundStates}, and preliminary searches have already extended sensitivity to unexplored parameter space~\cite{Witte:2022cjj,MainGapPaper,Battye:2023oac}.

Recent work in this field has demonstrated that the quasi-periodic 
vacuum electromagnetic fields in the polar caps of neutron stars can give rise to an enormous injection of axions (i.e., axion field amplitude) with energies (i.e., frequencies) $\omega_a \sim \mathcal{O}({\rm MHz - GHz})$~\cite{Prabhu:2021zve,MainGapPaper}. Should the axion mass $m_a$ lie roughly between ($10^{-9} {\rm eV} \lesssim m_a \lesssim 10^{-4} \, {\rm eV}$)\footnote{Unit convention: $c=\hbar = \epsilon_0 = 1$, and $e=\sqrt{4\pi\alpha}$.}, a sizable fraction of the produced axions will be  gravitationally bound to the neutron star.\footnote{An axion field configuration is a coherent state of many axion particle quanta. Here, we use both field and particle descriptions.} As these are feebly-interacting particles, the energy stored in gravitationally-bound axions cannot be easily dissipated, and accumulates over the lifetime of the pulsar, generating enormous axion densities~\cite{BoundStates}.
In this letter, we study the interplay between the growing axion cloud and the electromagnetic fields sourced by the pulsar itself, identifying striking signatures arising from axions in these systems. We focus specifically on the dynamics taking place in the polar caps of pulsars, as they contain large axion field values~\cite{BoundStates} and pockets of low plasma density that allow for the effect of axion-induced electric fields to be particularly pronounced.

\begin{figure}
    \centering
    \includegraphics[width = 0.48\textwidth]{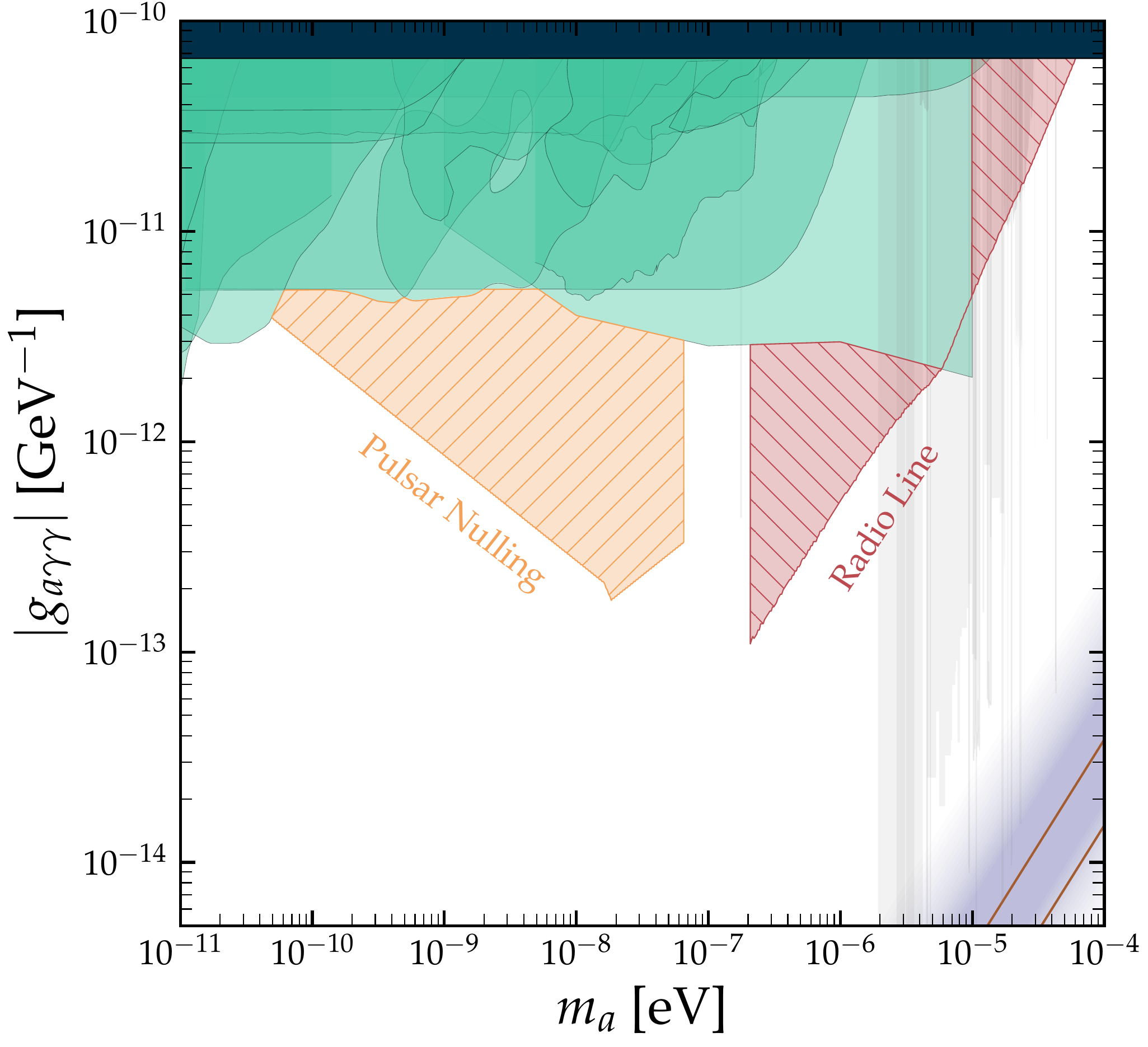}
    \caption{\label{fig:limit} Parameter space in which bound axions could induce a periodic nulling of (orange), or an observable radio line from (red), the three pulsars studied in this work. Shown for reference is the QCD axion band (purple), constraints from indirect astrophysical searches~\cite{Wouters:2013hua,HESS:2013udx,Payez_2015,Fermi-LAT:2016nkz,Meyer:2016wrm,Marsh:2017yvc,Reynolds:2019uqt,Xiao:2020pra,Li:2020pcn,Dessert:2020lil,Dessert1_2022,Dessert2_2022,MainGapPaper} (teal), CAST~\cite{Anastassopoulos2017} (dark blue), axion haloscopes~\cite{Sikivie1983,DePanfilis:1987dk,Hagmann:1990,Hagmann:1998cb,Asztalos:2001tf,Asztalos:2009yp,Du:2018uak,Braine2020,Bradley2003,Bradley2004,Shokair2014,HAYSTAC,Zhong2018,Backes_2021,mcallister2017organ,QUAX:2020adt,Choi:2020wyr,CAST:2020rlf} (grey; assumes axions are dark matter), and radio line searches in neutron stars~\cite{Foster:2022fxn,Battye:2023oac} (grey; assumes axions are dark matter) .}
\end{figure}

The primary findings of this work are twofold. First, we demonstrate that for sufficiently small axion masses, the axion field will induce a periodic screening of the electric field, suppressing particle acceleration and pair production, and leading to a quasi-periodic cancellation of radio pulses. For heavier axions, the small-scale fluctuations of the axion field drive the axion-induced voltage drop across the polar cap to zero, washing out the impact of the axion. Nevertheless, in this mass range we show that during the so-called `open phase', \ie a short-lived phase that precedes the onset of pair production, axions can produce a spectral line with $\mathcal{O}(5)\%$-level width (set by the axion velocity dispersion) that may be observable using existing radio telescopes; owing to the variation in neutron star mass/radius ratios~\cite{Miller:2019cac,Miller:2021qha},  there should also exist frequency variations at the ${\cal O}(20)\%$ level across the neutron star population due to gravitational redshifting. We use three known pulsars to illustrate the potential sensitivity of such probes, showing that future observations may be able to significantly improve sensitivity to axions over a wide range of parameter space (see Fig.~\ref{fig:limit}).

\section{Polar cap dynamics in the presence of axions}

The generalization of Gauss' and Ampere's laws in the presence of axions are given by~\cite{Wilczek:1987mv}:
\begin{align}
\nabla \cdot \vec{E}& = \rho - g_{a \gamma \gamma} \vec{B} \cdot \nabla  a\, , \label{Gauss}\\
\nabla \times \vec{B} - \dot{\vec{E}} &= \vec{\jmath} +  g_{a \gamma \gamma}\dot{a} \vec{B} + g_{a \gamma \gamma} \nabla a \times \vec{E}  \, ,
\label{Ampere}
\end{align}
which suggests the axion field $a$ produces effective charge and current densities $\rho_{a} = - g_{a\gamma\gamma} \vec{B} \cdot \nabla a$ and  $\vec{\jmath}_{a} = g_{a \gamma \gamma}\dot{a} \vec{B} + g_{a \gamma \gamma} \nabla a \times \vec{E}$. In most contexts, the corrections from the axion are sufficiently small to only perturbatively alter the electrodynamics; we show below, however, that this not necessarily true in the polar caps of pulsars.

Let us start with an overview of the conventional picture of electrodynamics in the polar caps (see \eg~\cite{RudermanSutherland1975,timokhin2010time,TimokhinArons2013} for further details). In the absence of a dense ambient plasma, the large-scale magnetic field together with the rotating, conducting, neutron star will induce a large electric field; the force exerted by the component of this electric field oriented along the magnetic field, $E_{||}$, can greatly exceed that of gravity, and can serve to directly extract charges from the stellar atmosphere. Along open magnetic field lines, this process drives an out-flowing current which serves to sustain the large-scale twist of the magnetic field, $j_m \equiv \nabla \times B$. 
When the current is super-Goldreich-Julian (GJ), $j_m > \GJ$, where $\GJ = -2 \, \vec{\Omega} \cdot \vec{B}$ (with $\vec{\Omega}$ and $\vec{B}$ being the angular velocity and magnetic field of the pulsar), the system develops a large voltage drop, accelerating electrons, which in turn generate high-energy curvature photons that initiate $e^\pm$ production. The resulting dense plasma dynamically screens the electric field via a non-linear quasi-periodic process that has been shown to source coherent radio emission (see \eg~\cite{Philippov:2020jxu}). If instead the supplied current is  sub-GJ (\ie $j < \GJ$), the extracted charges screen the field, and the system instead tends toward a steady state in which acceleration, pair production, and coherent radio emission are suppressed.

The stark difference in behavior observed between super- and sub-GJ current densities can be illustrated using a simplified one-dimensional evolutionary model~\cite{TimokhinArons2013}.
This toy model is expected to be a reasonable approximation to the initial behavior of the open phase of the gap (particularly for young, active, pulsars). Here, we generalize this to include the impact of an axion background, showing that axion charge densities $\rho_a \gtrsim \mathcal{O}(10\%)  \, \times \, \GJ$ can profoundly alter the dynamics of these systems. 
The one-dimensional model follows the acceleration of the primary particles,
which to a very good approximation 
propagate along the magnetic field lines. The electromagnetic field is assumed to be that of a global force-free magnetosphere with 
a constant charge density $\GJ$ and current density $\alpha\GJ$, determined by the (rotating) boundary conditions, plus an additional electric field parallel to the magnetic field and corresponding charge density which are present because the supply of charge from the neutron star atmosphere does not match that of the force-free configuration. 
Using \eqref{Gauss}, \eqref{Ampere}, and the Lorentz force law, and assuming a single sign for the charge (a good approximation when starting with a vacuum field at the polar cap), one can solve for the charge density and velocity (or Lorentz factor $\gamma(s)$) and the electric field 
component $E_s$ as functions of 
distance $s$ along the field lines, with the boundary conditions $\gamma(0)=1, \gamma_{,s}(0)=0$ at the surface of the star, $s=0$. This yields
\begin{equation}\label{eq:gamm_evol}
  \frac12\gamma_{,s}^2  =  \left(1-\frac{\rho_a}{\GJ}\right)(1-\gamma) +\left(\alpha-\frac{j_a}{\GJ}\right)\sqrt{\gamma^2 -1}\,,
\end{equation}
with $s$ in units of the Debye length scale, $\lambda_D = 1 / \omega_{p, {\rm GJ}}$.

\begin{figure}
    \centering
    \includegraphics[width=0.49\textwidth, trim={0cm .8cm .4cm 1.6cm}, clip]{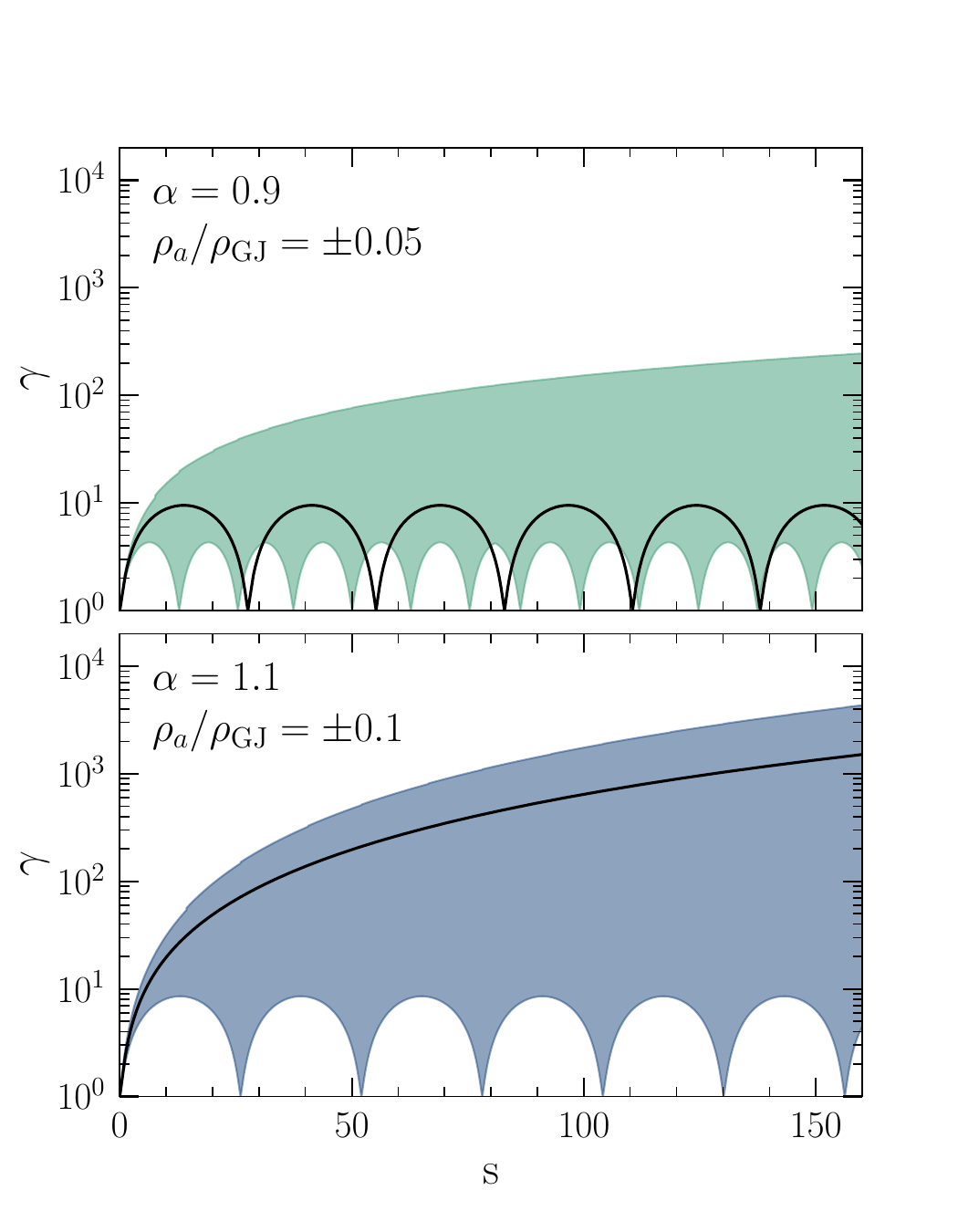}
    \caption{The evolution of the primary electron beam Lorentz factor (\ie the solution to Eq.~\ref{eq:gamm_evol}) as a function of distance $s$ 
    from the neutron star surface (in units of $\lambda_D$) for different initial discharge parameters $\alpha$. The standard case without axions is shown in black for $\alpha = 0.9$ 
    (top) and $1.1$ (bottom), and the effect of perturbing the system with a constant axion charge density $\rho_a$ is highlighted in the colored bands.}
    \label{fig:oneD}
\end{figure}

The ratio $\alpha \equiv j_m/\GJ$ is called the {\it discharge parameter}, 
since it governs the screening of the electric field. We show in Fig.~\ref{fig:oneD} the evolution of the electron Lorentz factor without axions (black) for $\alpha = 0.9$ (top panel)  and $\alpha = 1.1$ (bottom panel). For $\alpha > 1$, the energy continues to increase as the primary particles move away from the surface. As these particles are accelerated they emit curvature radiation along the field line; when the primary particles reach boost factors of $\gamma \sim 10^6-10^8$, the gamma-rays begin pair producing in the background magnetic field, and the production of these pairs leads to a screening of the ambient  electric field~\cite{Sturrock1971a, RudermanSutherland1975,Arons:1979bc,TimokhinHarding2015}. The evolution of the dynamical screening phase is highly non-linear, and understanding how axions affect this screening (and the subsequent generation of coherent radio emission) requires dedicated  simulations; nevertheless, for the purposes of this letter it is sufficient to focus on pair production as being a necessary step in this process. For  $\alpha < 1$\footnote{We note that pair production also occurs when $\alpha < 0$. These regions correspond to field lines supporting return currents, which serve to carry charges produced at large radii back to the star. For simplicity, we focus here only on the case of $\alpha > 0$.  }, on the other hand, the evolution of the primary current density is dramatically different. Here, the primary particles immediately begin screening the electric field; consequently, these particles never reach large Lorentz factors, and pair production never occurs.  In general, $\alpha$ is a spatially-varying function over the footprints of the open field lines, typically taking on $\alpha \sim \mathcal{O}(1)$ values, and with the characteristic variance on the order of  $\mathcal{O}(10\%)$~\cite{Gralla:2016fix,Gralla:2017nbw}.

The impact of axions on the evolution of the primary particles is shown in Fig.~\ref{fig:oneD}.
We assume the axion charge and current densities are oscillatory functions (as expected for classical fields)
which are constant over the time and length scales corresponding to
the evolution in Fig.~\ref{fig:oneD}, and we set $j_a = 3 \rho_a \sim \frac{\omega}{k} \rho_a$, which is a natural value of axions gravitationally bound to the neutron star -- see Sec.~\ref{Sec:AxionProd}. 
The shaded regions encompass the maximal and minimal $\gamma$ at each point along the trajectory,  
for values of $\rho_a$ falling between the indicated limits.
For sufficiently large $\rho_a$, the presence of the dense axion field induces transitions between the qualitatively different regimes\footnote{In a fully kinetic calculation, the stop-and-go electron flow configuration leads to an instability and development of a trapped particle population \cite{TimokhinArons2013}. Future self-consistent simulations will show how the dynamical evolution is  modified in the presence of substantial axion background.}. When the axion coherently suppresses particle acceleration along not just one, but a sufficiently large bundle, of open magnetic field lines, one expects a coincident suppression of coherent radio emission -- this is a novel signature which could provide striking evidence of axions in these systems.\footnote{Conversely, if the axion density could remain high enough when a pulsar nears its death, it might 
prolong the radio emission.}

Thus far we have shown that the behavior of pair production can be dramatically altered by an axion field with $\rho_a \gtrsim 0.1 \times \GJ$ 
-- in the following section we demonstrate that such field values can naturally be reached in many systems. Note that  we do not address the impact of axions on the non-linear dynamical screening phase. Instead, we focus on the fact that coherent radio emission requires pair production, which in turn requires high-energy particles. Future work will develop simulations of axion electrodynamics extending to the non-linear phase.

\section{Axion production in polar caps and pulsar nulling}\label{Sec:AxionProd}

We now turn our attention to the question of how, and when, large axion field values near the surfaces of neutron stars are expected to arise.

It has recently been shown that oscillations induced by the dynamical screening of the electric field in the polar cap source axions with characteristic energies $ \omega_a \lesssim 10^{-4}$ eV (this is a consequence of the fact that $g_{a\gamma\gamma}\vec{E} \cdot \vec{B}$ enters as a source term in the axion's equation of motion). If the axion mass lies in the range $10^{-10} \lesssim m_a \lesssim 10^{-4}$ eV, a sizable fraction of the produced axions will be, at most, semi-relativistic, and will remain gravitationally bound to the star itself. Furthermore, since axions interact feebly with the ambient matter they naturally accumulate on long, astrophysical, timescales. For axions in this mass range one  thus  concludes that  pulsars are generically surrounded by dense clouds of axions. That said, is the density of these clouds high enough to impact the electrodynamics as described in the previous section? The answer will be 
affirmative provided axion production is large and energy dissipation (extracting energy from the cloud) remains small. For simplicity we focus here on the case of light axions (with $\lambda_a \gtrsim h_{\rm gap}$), showing that both conditions can be satisfied for a wide range of parameter space. In the SM we derive similar conclusions for the case of heavy axions ($\lambda_a \lesssim h_{\rm gap}$).

The production of axions in polar caps has been studied using both a semi-analytic model and a kinetic plasma simulation~\cite{MainGapPaper}. Here, we use the result of~\cite{MainGapPaper} to derive analytic estimates of the energy injection rate and the axion density in the polar caps.

The rate at which axions extract energy from the electromagnetic fields is approximately given by~\cite{Prabhu:2021zve,MainGapPaper}
\begin{eqnarray}\label{eq:inj_approx}
\frac{dE_{\rm inj}}{dt} &\sim&  \zeta^2 \frac{g_{a\gamma\gamma}^2 \, m_a^3}{81 \times \tau} \, r_{\rm pc}^4 \, h_{\rm gap}^4 \, B_0^4 \, \Omega^2 \,,
\end{eqnarray}
where $r_{\rm pc} \simeq R_{\rm NS} \sqrt{R_{\rm NS} \, \Omega}$ is the polar cap radius, $R_{\rm NS}$ is the radius of the neutron star, $\Omega$ its rotational frequency, $B_0$ the surface magnetic field strength, $\tau \sim 10 \, h_{\rm gap}$ is the gap periodicity, and 
\begin{eqnarray}\label{eq:height}
   h_{\rm gap} \sim 12 \, {\rm m} \, \left( \frac{R_c}{10^7 \, {\rm cm}} \right)^{2/7}  \tilde{\Omega}^{-3/7}  \, \tilde{B}_0^{-4/7}  \, 
\end{eqnarray}
is the gap height. In Eq.~\ref{eq:height}, we have introduced the characteristic radius of curvature of open field lines $R_c \simeq 9 \times 10^7 {\rm cm} \, \sqrt{ P  / (1 \, {\rm sec}) } $~\cite{TimokhinHarding2015,Timokhin:2018vdn}, with $P = 2\pi/\Omega$, 
and used the $\tilde{X}$ notation to denote quantities $X$ which are normalized with respect to those of the Crab pulsar~\cite{Crab93, CrabMoreRecent} (\eg, $\tilde{\Omega} \equiv \Omega / \Omega_{\rm crab}$). Finally, we have also introduced an order-one fudge factor $\zeta$, characterizing the size of $E_{||}$ with respect to the maximal value $E_{\rm max} \sim \rho_{GJ} \times  h_{\rm gap}$, with $\GJ \sim -2 \vec{B} \cdot \vec{\Omega}$ being the Goldreich-Julian (GJ) charge density (\ie $E_{||} \equiv \zeta E_{\rm max}$) ~\cite{Goldreich:1969sb,RudermanSutherland1975}.
 
Assuming the cloud grows linearly on long timescales (compared to the variability of $\vec E\cdot\vec B$), the axion energy density at the surface of the neutron star after a time $t$ is roughly given by \small
\begin{multline}\label{Eq:DensityTime}
\epsilon_0(t) \sim   \frac{3 \, t \,}{16 \pi \,r_{\rm NS}^3}\frac{dE_{\rm inj}}{dt}  
    \sim 5 \times  10^{23} \frac{{\rm GeV}}{\rm cm^{3}}   g_{-12}^{2}  m_{-8}^3 \, \tilde{t}_{\rm age} 
     (\tilde{B}_0\tilde{\Omega})^{8/7},
\end{multline}
\normalsize
where $g_{X} = g_{a\gamma\gamma} / 10^{X} \, {\rm GeV^{-1}}$ and $m_{X} = m_{a} / 10^{X} \, {\rm eV}$, and the exponent ${{8/7}}$ comes from the implicit dependence on  $B_0$ and $\Omega$ entering $h_{\rm gap}$ and  $r_{\rm pc}$.
We can now determine whether sufficiently large axion densities are achievable in these systems by comparing Eq.~\ref{Eq:DensityTime} with the back-reaction density, defined as the energy density for which $\rho_a = \xi \GJ$ (with $\xi \sim \mathcal{O}(1)$ an order-one pre-factor). Given an axion field energy density $\frac12 m_a^2 a^2$ and the axionic charge density $\rho_a\sim g_{a\gamma\gamma}Bk_a a$ (\eqref{Gauss}), we have
 \begin{eqnarray}\label{eq:rho_br} 
\epsilon_{\rm br} &\sim& \xi^2  \frac{2 \Omega^2 m_a^2}{g_{a\gamma\gamma}^2 k_a^2 }\, 
\sim 5 \times 10^{20} \, \frac{ {\rm GeV}}{ {\rm cm^3}} \,\tilde{\Omega}^2 \, \left( \frac{1}{g_{-12}} \, \frac{\xi}{0.1}\right)^2
\end{eqnarray}
Comparing this with Eq.~\ref{Eq:DensityTime}, we see that back-reaction densities are  achievable in many systems provided 
$g_{-12}\gtrsim \mathcal{O}(0.1)$.

Next, one must determine whether there exists an energy dissipation mechanism that could quench the growth of the cloud before reaching  $\epsilon_{\rm br}$. For light axions, on-shell photon production is kinematically blocked by the presence of the plasma, and thus all energy dissipated from the axion field must proceed indirectly through the local current -- that is to say, the axion-induced electric field helps to accelerate local charges, which in turn dissipate energy via radiation~\cite{harding1998particle}. The net (time-averaged) energy loss can be obtained by computing $\left< \vec{j} \cdot \vec{E}_a \right>$, where $E_a$ is the axion-induced electric field and $j_e$ the current. 
In the SM~\footnote{There we make use of Ref.~\cite{Raffelt:1996wa, Beutter:2018xfx, Jackson, Caputo:2021efm} to perform our computations.} we show that the leading-order contribution to the axion-induced electric field is given by 
\begin{equation}\label{eq:inducedE}
    \vec{E}_a \sim g_{a\gamma\gamma}  \frac{\sqrt{2 \, \epsilon_a}}{m_a} \vec{B} \, e^{-i m_a t} \times \mathcal{F}(m_a, h_{\rm gap}, \omega_p) \, ,
\end{equation}
where the factor $\mathcal{F}$ captures the in-medium suppression, with a uniform vacuum yielding $\mathcal{F} = 1$, and a dense uniform plasma yielding $\mathcal{F}\sim (m_a / \omega_{p, {\rm eff}})^2 \ll 1$ (where $\omega_{p, {\rm eff}} \equiv \left<\omega^2_p / \gamma^{3}\right>^{1/2}$, with $\left<\, \right>$ denoting the average over the distribution functions and the species present in the plasma, and the plasma frequency of a single species $s$ defined as $\omega_{p, s}^2 \equiv q_s^2 n_s/ m_s$, with $q_s, n_s,$ and $m_s$ the charge, number density, and mass~\cite{arons1986wave}) 
(with non-uniform geometries, as is relevant for this case, one obtains a 
suppression interpolating between these limiting regimes, see SM). One can see from Eq.~\ref{eq:inducedE} that the electric field is heavily suppressed once plasma is produced, and thus the energy dissipation will be most efficient during the open phase of the gap. 
Equating the axion production and dissipation rates, one can estimate the maximal axion energy density (\ie the saturation density) at the neutron star surface as
\begin{eqnarray}\label{eq:saturate}
\epsilon_{\rm sat}   &
   \sim &  8 \times 10^{19}  \, \frac{\rm GeV}{\rm cm^3} \, \zeta^2  \, m_{-8}^3 \, \tilde{B}_0^{9/7} \, \tilde{\Omega}^{9/7} \, .
\end{eqnarray}
Pulsar nulling requires efficient axion production, $\epsilon_0/\epsilon_{\rm br} \propto B^{8/7}\Omega^{9/7}\gtrsim 1$, and inefficient energy dissipation, $\epsilon_{\rm br}/\epsilon_{\rm sat} \propto B^{-9/7}\Omega^{5/7} \lesssim 1$. These conditions can be accommodated for pulsars with large magnetic fields and large periods. 

In Fig.~\ref{fig:limit} we highlight the parameter space for which the axion can dramatically impact the polar cap dynamics. For sufficiently small axion masses (with $\lambda_a \gg h_{\rm gap}$), the axion cloud will uniformly suppress acceleration (and thus pair production and dynamical screening) across the polar cap, leading to a short-lived nulling of the radio signal (see shaded region in Fig.~\ref{fig:limit} labeled `Pulsar Nulling'). Since axions are on-shell, any nulling is only temporary, implying the suppression of the radio emission should be approximately periodic with the characteristic timescale set by the axion energy. This implies a well-defined periodic structure in the radio emission on timescales $t \sim \mathcal{O}({\rm ns - \mu s})$, which is observable using current telescopes (see \eg \cite{hankins2003nanosecond} for an example of high time-resolution observations).

\vspace{-.2cm}
\section{Incoherent dipole emission}
\vspace{-.2cm}

The previous section focused on the case where axions could be treated as being coherently in-phase across the gap, which is only valid for sufficiently light axions. There are two primary differences that arise when one considers the implications of heavier axions (with $\lambda_a \ll h_{\rm gap}$). First, high-mass axions are produced by rapid small-scale fluctuations in the plasma -- since the amplitude of these fluctuations are much smaller than the vacuum field, the production rate is suppressed relative to the low-mass limit. Second, the axion field is now incoherent over the gap, meaning one must account for the stochasticity of the axion phase; the incoherence heavily suppresses the effect of the axion on the electrodynamics, and in this regime the existence of local densities $\epsilon \sim \epsilon_{\rm br}$ may not guarantee the existence of pulsar nulling (see SM where we adopt the approach of Ref.~\cite{Tolman:2022unu}). Nevertheless, the oscillation timescale of the axion field is now sufficiently fast to produce on-shell electromagnetic radiation during the open phase of the gap. In particular, we find that the axion field can produce a narrow, highly beamed, radio line. The origin of this line stems from the fact that an oscillating axion in a background magnetic field behaves equivalently to an oscillating dipole moment. The radiation produced from the oscillating dipole is highly incoherent in the limit $\lambda_a \gg h_{\rm gap}$, and is thus heavily suppressed at large masses; nevertheless, the total emission is sufficiently strong to give rise to observable signatures. Assuming that the axion coupling is sufficiently large that the back-reaction density is reached, \ie $g_{a\gamma\gamma} \geq g_{P}^{\rm min}$ where $g_{P}^{\rm min}$ is the minimum coupling required for pulsar $P$ to produce an axion cloud with density $\epsilon_{\rm br}$, and that beaming~\footnote{Beaming is a consequence of the dense plasma layers surround the boundary of the polar cap, which serve as mirrors, confining the radiation (see SM).} confines the emission to an angular opening of the size of the polar cap (\ie $\theta_{\rm max} \sim  r_{pc} / R_{\rm NS}$), 
we infer (see the SM) an observable flux density of
\small
\begin{eqnarray}\label{eq:sdense}
    S_\gamma (g_{a\gamma\gamma} &\geq &g_{P}^{\rm min}) \equiv \frac{dE_\gamma/dt}{d\Omega \, \mathcal{B}  d^2 \, }  \sim 38 {\rm \, Jy} \Big(\frac{\xi}{0.1}  \frac{2 \, {\rm{kpc}}}{d}  \Big)^2 \frac{\tilde{B}_0^{-2/7} \tilde{\Omega}^{-9/7 }}{m_{-6}} , \nonumber
\end{eqnarray}
\normalsize
where we set the bandwidth to be $\mathcal{B} = m_a/50$ (note that this includes a suppression from the incoherence).  For reasonable parameters this flux can greatly exceed the intrinsic radio flux of the pulsar. Finally, we note that at couplings $g_{a\gamma\gamma} \leq g_{P}^{\rm min}$ the flux density scales as $S_\gamma \propto g_{a\gamma\gamma}^4$, and thus this signal is strongest for pulsars reaching the back-reaction density. \\ \vspace{.5cm}

\section{Results and Conclusions } \label{sec:results}
\vspace{.2cm}

Throughout this work we have chosen to use the well-studied Crab pulsar as a reference. There exist pulsars, however, that are more favorable targets to observe both pulsar nulling and non-resonant radio emission.
We thus use the ATNF catalogue to identify better targets for both observables~\cite{jankowski2019utmost, bell2016time, jankowski2018spectral, weltevrede2011glitch}. In doing the selection, we enforce that the inferred magnetic field remain below the Schwinger field strength, the pulsar is not a millisecond pulsar, and the spin-down rate is $\dot{E} > 10^{34}$ erg/s (the latter condition being a rough estimator for when pair cascades are one-dimensional~\cite{SashaReview}). Among the optimal candidates for pulsar nulling is J1119-6127, while the preferred pulsar for radio emission is B1055-52; the parameters for each pulsar are given in Tab.~I of the SM. In the SM we also provide a discussion about pulsar magnetic field evolution~\cite{Spitkovsky2006, Philippov2014, lyne2013evolution, lyne201545, rookyard2015constraints}, and identify the parameter space for which back reaction, pulsar nulling, and observable radio emission would arise for each of the three pulsars. These results are combined in Fig.~\ref{fig:limit} in order to provide  an idea of the axion parameter space that could be explored using future observations.

This work is an initial investigation into previously unknown phenomena, providing rough estimates illustrating when axion back-reaction can occur, and identifying what might be the most robust associated
signatures. Fig.~\ref{fig:limit} shows that both pulsar nulling and non-resonant axion-induced radio emission from the polar cap region may provide sensitive probes of axions across a wide range of unexplored parameter space. The complexity of axion-electrodynamics in these environments necessitates  follow-ups using state-of-the-art numerical simulations (see \eg~\cite{Philippov:2020jxu,Cruz:2021hku}) that include strong axion back-reaction; this highly non-linear regime might produce striking new signatures, such as a persistent quasi-periodic suppression of the radio emission and shifts in the neutron star death line.

\vspace{-.5cm}
\section{Acknowledgments}%
 The authors would like to thank Georg Raffelt, Edoardo Vitagliano, Jamie McDonald, Ani Prabhu, Thomas Schwetz, Dion Noordhuis, Andrew Shaw, and Cole Miller for their useful discussions.
 AC acknowledges the hospitality of the Flatiron Center for Computational Astrophysics during the initial stages of this project. SJW acknowledges support from a Royal Society University Research Fellowship (URF-R1-231065), and through the program
Ram\'{o}n y Cajal (RYC2021-030893-I) of the Spanish
Ministry of Science and Innovation. This article/publication is based upon work from COST Action COSMIC WISPers CA21106, supported by COST (European Cooperation in Science and Technology). This work was supported by a grant from the Simons Foundation (MP-SCMPS-00001470) to AP.
The work of TJ was supported in part by NSF grant PHY-2309634.

\vspace{6cm}

\onecolumngrid
\begin{center}
  \textbf{\large Supplementary Material for Axion-Induced Pulsar Nulling and Non-Resonant Radio Emission}\\[.2cm]
  \vspace{0.05in}
  { Andrea Caputo, Samuel J. Witte, Alexander Philippov, Ted Jacobson}
\end{center}

\twocolumngrid
Below, we provide additional details justifying some of the approximations, and outlining the derivations of various quantities quoted in the main text. These include: a look at the velocity distribution of the axion cloud near the neutron star surface, a one-dimensional analysis of the discharge parameter, the derivation of the axion-induced electric field, a derivation of the axion production rate, back-reaction density, and energy dissipation rate (both at high and low axion masses), a look at the impact of magneto-rotational spin-down, details on the sensitivity analysis, and a discussion on possible resonant transitions during the phase of pair production. 

\section{Velocity distribution near the neutron star}

\begin{figure}[t!]
    \centering
    \includegraphics[width = 0.45\textwidth, trim={0cm 0cm 0cm 0cm}, clip]{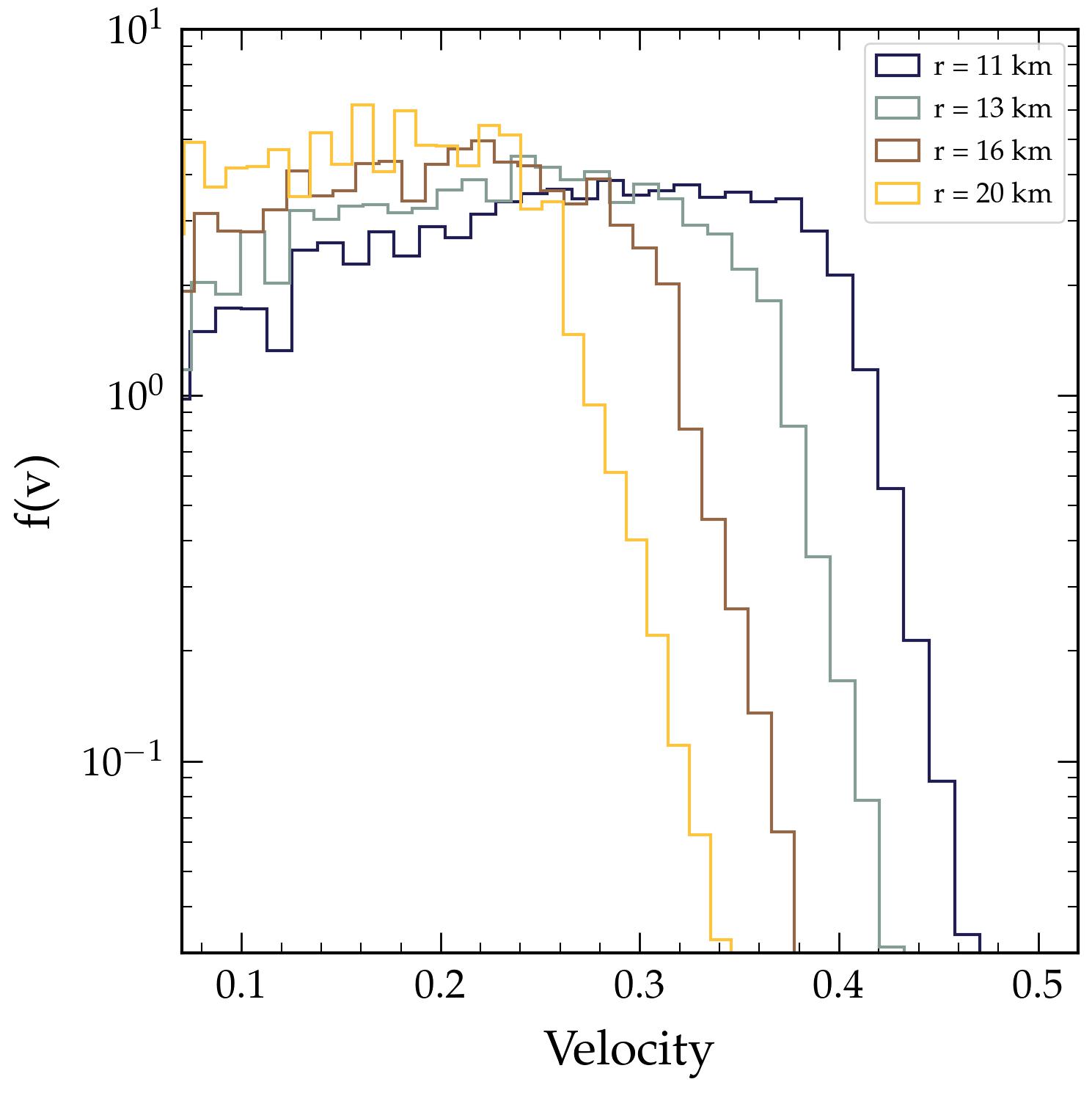}
    \caption{Normalized axion velocity distribution $f(v)$ as a function of velocity evaluated at four different radii, averaged over the all angles. This distribution is used to justify  the approximation $k_a \sim m_a / 3$ near the neutron star surface.  }
    \label{fig:phaseS}
\end{figure}

Throughout the main text we have used the approximation that the axion phase space in the gap itself is dominated by characteristic axion momenta $k_a \sim m_a / 3$. In order to justify this approximation we compute the initial axion spectrum for the Crab pulsar, sample momenta from the derived production rate (in units of number of axions per unit k-mode), and follow the evolution of these sampled axion bound states for a timescale on the order of $t \sim 1 \, \rm s$ (corresponding to $3\times10^3$ neutron star crossing times). These trajectories are traced following treatment of~\cite{BoundStates}. From the sampled trajectories, we draw a spherical surface at some radius $r$ around the neutron star and determine the crossing points of each of the trajectories. Each crossing point is re-weighted using the procedure outlined in~\cite{BoundStates}. We plot the normalized distribution of weighted samples in Fig.~\ref{fig:phaseS}.

\section{Axion-induced Electric Field}

Here, we provide additional details outlining the derivation of the induced electric field and the energy dissipation mechanisms. We begin by following Ref.~\cite{Beutter:2018xfx} in identifying the transition amplitude from an electron between initial $\left. |i \right>$ and final $\left. |f \right>$ states at zeroth order in axion-photon coupling:
\begin{equation}
    \mathcal{A} = -i \left< f| \int d^4x \mathcal{H}_I(x) | i \right> = -i \int d^4x \, J^\mu(x) A_{\mu}^{\rm ext}(x) \, ,
\end{equation}
where we have defined the vector potential induced by an external electromagnetic field, and we have defined $J^\mu \equiv e \left<f| \bar{\psi} \gamma^\mu \psi | i \right>$. We can determine the analogous expression at second order from the axion-induced electric field as
\begin{equation}
\mathcal{A} = \frac{(-i)^2}{2!} \left< f|T \int d^4x \, \int d^4y \mathcal{H}_I(x) \mathcal{H}_I(y) | i \right>
\end{equation}
were $T$ is the time-ordered product, and $H_I$ includes the axion and fermion interaction terms. This amplitude can be re-expressed as 
\begin{equation}
    \mathcal{A} = -i \int d^4x J^\mu(x) A_{\mu}^{\rm ind}(x),
\end{equation}
where the induced vector potential is given by
\begin{equation} \label{eq:a_ind}
    A_{\mu}^{\rm ind}(x) = i g_{a\gamma\gamma}\int d^4y D_{\mu\nu}(x-y) \partial_\rho a(y) \tilde{F}^{\rho \nu}(y)
\end{equation}
with $D_{\mu\nu}$ being the photon propagator. At this point our derivation deviates from that in ~\cite{Beutter:2018xfx}, as we must consider the dressed propagator in medium, which is given (in the Feynman gauge) by 
\begin{equation}
    D_{\mu\nu}(x-y) = \int \frac{d^4q}{(2\pi)^4}\frac{-ig_{\mu\nu}}{q^2 - \Pi_{L,T}} \, e^{- i q (x-y)} \, ,
\end{equation}
with $\Pi_{L,T}$ the longitudinal and transverse polarization tensors. For a non-thermal plasma in the limit $q \rightarrow 0$, these are given by $\Pi_{L,T} \simeq \omega_p^2$~\cite{Jackson:1998nia,Raffelt:1996wa}, with $\omega_p$ the plasma frequency. In the limit that $q^2 \sim m_a^2 \ll \omega_p^2$, the $d^4q$ integration just gives $(2\pi)^4 \delta^{(4)}(x-y) / \omega_p^2$, so the induced vector potential simplifies to
\begin{equation}
    A_{\mu}^{\rm ind}(x) = i g_{a\gamma\gamma} \frac{\partial^\rho a(x) \tilde{F}_{\rho \mu}(x)}{\omega_p^2} \, .
\end{equation}
Note that this `contact interaction' approximation is valid so long as $k \sim m_a \ll \omega_p$, which is valid for most of the magnetosphere of the Crab pulsar, so long as $m_a \lesssim 10^{-5}$ eV. Taking the non-relativistic limit we see that the dominant contribution to the electromagnetic field is given by
\begin{equation}\label{eq:EInd}
    \vec{E}^{\rm ind} = -\partial_t \vec{A}^{\rm ind} \simeq i g_{a\gamma\gamma} \Big(\frac{m_a}{\omega_p}\Big)^2  \frac{\sqrt{2 \,\epsilon_a(r)}}{m_a} \vec{B} \, e^{i (k_a \cdot r - m_a t)},
\end{equation}
where we recognize the typical in-medium suppression factor $(m_a/\omega_{\rm pl})^2$ (see \eg \cite{Caputo:2021efm} for a comparison in the context of dark photons). We can also compute the analogous expression for the induced magnetic field, which is given by
\begin{eqnarray}\label{eq:EInd}
    \vec{B}^{\rm ind} &=& -\nabla \times \vec{A}^{\rm ind} \\  &\sim& i g_{a\gamma\gamma}  \Big(\frac{k_a}{\omega_p}\Big)^2  \frac{\sqrt{2 \, \epsilon_a(r)}}{m_a}  E_{||} \, e^{i (k_a \cdot r - m_a t)} (\hat{x} + \hat{y}) \nonumber
\end{eqnarray}
where we have assumed $\nabla \times \vec{B} \simeq 0$ and $\vec{E} \simeq E_{||} \hat{z}$ in the polar cap, and $\partial_{x,y} (a E_{||}) \sim k a E_{||}$.  

The in-medium effect leads to a powerful suppression of the induced electric and magnetic fields when $m_a \ll \omega_p$. However, during the `open phase' of the gap has an effective plasma frequency on the order $\omega_p / \gamma^{3/2} \lesssim 10^{-12}$ eV, and thus we expect the dominant energy losses to arise during this short-lived phase. 

In order to analyze the behavior of the induced electric field during the open phase of the gap, we assume one can treat the gap itself as being pure vacuum surrounded by a sharp boundary of dense plasma; while this is slightly over-simplified, it allows us to obtain rough estimates of the induced electric field and the effect of this field on the system. Re-solving Eq.~\ref{eq:a_ind} with this geometry  we find 
\begin{equation}\label{eq:EInd}
    \vec{E}^{\rm ind} \simeq i g_{a\gamma\gamma} \mathcal{F}(m_a, h_{\rm gap}, \omega_p)  \frac{\sqrt{2 \,\epsilon_a(r)}}{m_a} \vec{B} \, e^{i (k_a \cdot r - m_a t)} \, ,
\end{equation} 
where, assuming $\omega_p \gg m_a$, one finds the suppression factor to be
\begin{eqnarray}
    \mathcal{F} \sim \begin{cases}
    1 \hspace{.05cm } & \lambda_a \ll h_{\rm gap}, r_{\rm pc}, R_{\rm NS} \\
    \left(\frac{h_{\rm gap}}{\lambda_a} \right) \hspace{.05cm } & h_{\rm gap} \ll \lambda_a \ll r_{\rm pc}, R_{\rm NS} \\
    \left(\frac{h_{\rm gap} r_{\rm pc}^2}{\lambda_a^3} \right)  & h_{\rm gap}, r_{\rm pc} \ll \lambda_a  \ll R_{\rm NS} \\
    \end{cases} \, .
\end{eqnarray}

\section{The Low Axion Mass Regime}

In this section we provide extra details about the derivation of the axion production rate, the back-reaction density derivation and the energy loss in the small mass regime, \textit{i.e} $\lambda_a \gg h_{\rm gap}$.

\subsubsection{Axion Production}

We start by outlining the derivation of the energy injection rate and the density of the axion cloud at the neutron star surface. Additional details and discussion can be found in~\cite{MainGapPaper}.

Starting from the axion's equation of motion, $(\Box + m_a^2)a(x) = g_{a\gamma\gamma} \vec{E}(x) \cdot \vec{B}(x) $, one can show that rate at which energy is injected into axions is given by~\cite{MainGapPaper}
\begin{equation}\label{eq:erg_scalar}
    \frac{dE_{\rm inj}}{dt} = \frac{1}{2 (2\pi)^3 \tau} \int d^3 k |\tilde{\mathcal{S}}(\vec{k})|^2 \, ,
\end{equation}
where we have defined the source term for axions
\begin{equation} \label{eq:j_axion}
    \tilde{\mathcal{S}}(\vec{k}) = -g_{a\gamma\gamma} \int d^4x \, e^{i k \cdot x} \,  \vec{E}(x) \cdot \vec{B}(x)  \,,
\end{equation}
and $\tau$ is defined to be the characteristic period of the gap collapse process. It is then clear that in order to compute the axion production rates one needs the Fourier transform (FT) of the source term $\vec{E}(x) \cdot \vec{B}(x)$. We also note that in defining the FT of the source, we have implicitly enforced the on-shell condition $\omega_a^2 = k_a^2 + m_a^2$. 

In order to make quantitative estimates one the needs to specify $\vec{E}(x) \cdot \vec{B}(x)$ and its temporal evolution. Within the simplifying assumptions outlined in the main text, and working in the limit $m_a / (2\pi) \ll \tau^{-1}$ (the opposing limit will be discussed below), the typical energy injection rate can be estimated using Eq.~\ref{eq:erg_scalar}, and is given by 
 \begin{eqnarray}\label{eq:inj_approx2}
\frac{dE_{\rm inj}}{dt} &\sim& \frac{g_{a\gamma\gamma}^2}{2(2\pi)^3}\underbrace{\frac{4\pi \, m_a^3}{81}}_{d^3k}\underbrace{\pi^2 \, r_{\rm pc}^4 h_{\rm gap}^2}_{V_{\rm gap}^2} \frac{\tau_{\rm open}^2}{\tau} \zeta^2 B_0^2 \underbrace{4\, B_0^2\Omega^2 h_{\rm gap}^2}_{E_{\rm max}^2} \nonumber \\  &\sim&  \zeta^2 \frac{g_{a\gamma\gamma}^2 \, m_a^3}{810} \, r_{\rm pc}^4 \, h_{\rm gap}^5 \, B_0^4 \, \Omega^2, 
\end{eqnarray}
where we have assumed $\tau_{\rm open} \sim h_{\rm gap}$ and $\tau = \tau_{\rm open} + \tau_{\rm closed} \sim 10 \, h_{\rm gap}$. Here $\tau_{\rm open}$ is the time scale of the ``open phase" (occurring prior to pair productio nwhen the gap is effectively vacuum) and  $\tau_{\rm closed}$ is the timescale on which the gap is screened. We have taken this latter to be $10$ times the light crossing time. In general, the exact value of $\tau_{\rm close}$ is not known -- it must be at least one light crossing time, and one-dimensional high-resolution simulations has suggested that it could be as large as $\sim 10$ light crossing times. We adopt the upper limit on this in order to remain maximally conservative in our prediction of the axion production rate.

The characteristic density at the surface is obtained by noting that the density profile outside the star scales as $\epsilon_a \propto r^{-4}$, and is roughly constant within the star. Using this profile, one can estimate the fraction of the energy being injected into bound states in radii $r \in [r_{\rm NS}, r_{\rm NS} + h_{\rm gap}]$; dividing off by the appropriate volume element and multiplying by the age of the star yields Eq.~6 in the main text.  

\subsubsection{Back-reaction density}

In order to compute the back-reaction density we need to evaluate the axion cloud effective charge density, $\rho_a = - g_{a\gamma\gamma} \vec{B} \cdot \nabla a$ and compare it with the GJ density, $\GJ$. Since particle acceleration and pair production is nearly one-dimensional, we are interested in comparing $\GJ$ with the value of $\rho_a$ averaged over the open phase of the gap and over a characteristic bundle of field lines. In the low mass regime ($\lambda_a \gg h_{\rm gap}$), the axion field is approximately uniform (both spatially and temporally during the open phase of the gap), and thus this trivially yields
\begin{eqnarray}\label{Eq:grad} 
\frac{1}{h_{\rm gap} \, \tau_{\rm open}} \int dt \int_{R_{\rm NS}}^{R_{\rm NS}+ h_{\rm gap}} & dz & \nabla a  \nonumber \\ = \frac{a(R_{\rm NS}+ h_{\rm gap}) - a(R_{\rm NS})}{h_{\rm gap}}\nonumber \\[2ex]
 \simeq   \frac{\sqrt{2\epsilon_a}}{m_a} k_a \, \cos(m_a \, \tau_{\rm open} + \delta),
\end{eqnarray}
where we introduced a random phase $\delta$ for the axion field. Imposing  $\rho_{a} \sim \xi \GJ$, where the factor $\xi \sim \mathcal{O}(0.1)$ has been introduced to account for the uncertainty in the back-reaction threshold, and using Eq.~\ref{Eq:grad} for the axion gradient, one finds Eq.~7 of the main text
 \begin{eqnarray}\label{eq:rho_brSM}
    \epsilon_{\rm br} &\sim& \xi^2  \frac{2 \Omega^2 m_a^2}{g_{a\gamma\gamma}^2 k_a^2 }\, \\
    & \sim& 5 \times 10^{20} \, \frac{ {\rm GeV}}{ {\rm cm^3}} \, \left( \frac{\Omega}{190.4 {\rm Hz}} \frac{10^{-12} \, {\rm GeV^{-1}}}{g_{a\gamma\gamma}} \frac{\xi}{0.1}\right)^2 \nonumber
\end{eqnarray}

\subsubsection{Energy dissipation rate}

Here we compute the energy loss of charges due to curvature radiation. We recall this is a crucial quantity to compute the saturation density (Eq.~9 in the main text), as curvature radiation provides an energy `sink' for the axion field. The rate of energy loss per particle due to curvature radiation is given by 
~\cite{harding1998particle}
\begin{equation}\label{eq:curveR}
    \dot{E}_{e^\pm} \simeq \frac{3 e^2}{2 R_c^2}\gamma^4 \, .
\end{equation}
Neglecting energy losses, the differential change in $\gamma$ as a function of distance $s$ is given by $d\gamma(s, \vec{E})/ds \sim e |\vec{E}| / m_e$. To compute the energy losses from the axion field, we can first compute the total energy dissipated across the gap in the form of curvature during a single vacuum phase 
\begin{eqnarray}\label{eq:e_diss}
    \frac{dE_{\rm dis}}{dt} &\sim& 
     \pi r_{\rm pc}^2 \, n_{\rm GJ} \, \int_0^{h_{\rm gap}} ds \,   \dot{E}_{e^\pm}(s)\, .
\end{eqnarray}
Writing the electric field as a bare component (assumed to be linearly proportional to the height) and a constant perturbation due to the axion, \ie $\vec{E}_{0}(s) \pm \vec{E}_a$ and therefore $\gamma \sim \frac{e}{m_e}\Big(\frac{E_0 s^2}{2\, h_{\rm gap}}+ E_a s\Big)$~\cite{harding1998particle}, one can expand Eq.~\ref{eq:e_diss} in powers of $\epsilon = |\vec{E}_a| / |\vec{E}_0|$. Since the first order term $\propto \vec{E}_a$ will induce a vanishing contribution when averaged on timescales $t \gg 2\pi/m_a$, the leading order contribution to the energy loss arises at $\mathcal{O}(\epsilon^2)$ -- time averaging this term (and working in the limit limit $h_{\rm gap }\ll \lambda_a \ll r_{\rm pc}$) we arrive to
\begin{eqnarray}\label{eq:e_diss_tavg2}
    \left<\frac{dE_{\rm dis}}{dt} \right> &\sim& 
     \frac{ g_{a\gamma\gamma}^2}{70\pi}  \, \epsilon_a \, B_0^5 \, \Omega^3 \, \frac{e^5  \, h_{\rm gap}^9 \, r_{\rm pc}^2}{m_e^4 \, R_c^2} \, .
\end{eqnarray}
Equating this expression to the energy injection rate in Eq.~4 leads to the saturation density in Eq.~9 of the main body. Analogous expressions can be derived for $\lambda_a \gg r_{\rm pc}, h_{\rm gap}$ by using the appropriate axion-induced electric field. Notice that we have included a factor of $\sim 1/10$ to account for the duty cycle of the open phase of the gap.

\section{The High Axion Mass Regime}
The main text of the manuscript has focused primarily on deriving quantities in the low-mass limit, as in this regime the axion field can be treated as coherent over the gap, implying the impact of the axion on the electrodynamics is greatly simplified. At large masses, this is no longer true -- the axion field stochastically samples random phases within the gap, implying many effects tend to cancel, or wash out. Here, we re-derive all the relevant quantities -- namely, the scaling of the axion production rate, the back-reaction density, and the rate of energy dissipation -- in the high mass limit.

\subsubsection{Axion production rate}

In general, the Fourier transform (FT) of $\vec{E}\cdot\vec{B}$ receives two contributions, one coming from the characteristic geometry of the gap, and the other coming from the small scale oscillations induced by the plasma screening process. Since oscillations can only exist on scales below the gap size, the second contribution is zero in the low mass limit (and as such has been neglected in the main text). At large axion masses, the volumetric contribution is expected to scale like $\tilde{\mathcal{S}} \propto (\lambda_a / L_{\rm gap})^3$, meaning the contribution to the axion production rate is dramatically suppressed at high masses. This effect is largely compensated, however, by the contribution arising from the small-scale oscillations in the plasma. In order to illustrate this effect, let us consider a 1-dimension toy example in which we consider a window function with a small scale oscillatory perturbation, \ie 
\begin{eqnarray}
f(x)  = & f_{\rm geometery} + f_{\rm oscillations} \\ =  & \, \Theta(x) \Theta(L-x)  + \epsilon \Theta(x) \Theta(L-x) \sin(k_{\rm osc} \, x), \, \nonumber
\end{eqnarray}
where $\Theta$ is the Heaviside step function, $L$ is the length of the window function, $\epsilon$ is the size of the oscillatory perturbation, and $k_{\rm osc}$ is the wavelength of the perturbation. In the limit $k_{\rm osc} \rightarrow 0$ we recover the purely geometric contribution which is merely the size of the gap ${\rm FT}(f )_{ k_{\rm osc} \rightarrow 0} \sim L$. In the alternative limit (and evaluating the FT for $k \rightarrow k_{\rm osc}$), one is instead dominated by the oscillatory contribution, finding  ${\rm FT}(f )_{k_{\rm osc} \rightarrow 0} \sim L \times (\epsilon / 2)$. 

Returning to the case of the polar cap, the amplitude of oscillations decreases as the oscillation frequency increases (\ie a denser plasma induces higher frequency damped oscillations in the electric field); that is to say, the normalization of the amplitude of oscillations in the previous example is in fact a function of the frequency itself, \ie $\epsilon(\omega)$. Using the analytic derivations of the one-dimension response derived in~\cite{Tolman:2022unu}, we can estimate that the amplitude of oscillations scales with the frequency approximately as $E_{||} (\omega)  \propto E_{||, 0} \times (\omega / \omega_0)^{-\xi}$, with $\xi \sim 0.5$. Working in one dimension, this suggests that at large axion masses (which for non-relativistic axions implies large oscillation frequencies) one expects the Fourier transform of $
\vec{E}\cdot\vec{B}$ to scale as $m_a^{-1/2}$; generalizing this result to three dimensions, and assuming the temporal integral in $\tilde{\mathcal{S}}$ follows similar trend to the spatial contributions\footnote{Note that it is somewhat unclear whether this approximation holds, however given that the dynamics of the system are operating in the relativistic regime, we expect some approximate consistency between the spatial and temporal evolution of the system. A detailed understanding of this re-scaling likely necessitates a more sophisticated numerical approach which is beyond the scope of this work.}, one finds that the high mass limit scales like
\begin{equation}\label{eq:einj_highm}
\Big(\frac{dE_{\rm inj}}{dt}\Big)_{\rm high} \sim \Big(\frac{dE_{\rm inj}}{dt}\Big)_{\rm low} \Big(\frac{1}{m_a \, L_{\rm gap}}\Big)^3 \Big(\frac{\omega_0}{m_a}\Big), 
\end{equation}
where $\Big(\frac{dE_{\rm inj}}{dt}\Big)_{\rm low}$ is defined in Eq.~4 of the main text. In order to derive a practical expression for the evolution of the density in the high mass limit we simply extract the functional dependence of Eq.~\ref{eq:einj_highm}, and ensure continuity at the transition mass with Eq.~4 of the main body. Using the parameters of the Crab, this leads to a density of
\begin{eqnarray}\label{Eq:DensityTimeHighMass}
\epsilon_0(t) \sim 10^{25} \, \frac{{\rm GeV}}{\rm cm^{3}} \, g_{-12}^{2} \, \frac{\tilde{t}_{\rm age}}{m_{-6}}  \,
     \tilde{B}_0^{24/7} \, \tilde{\Omega}^{24/7} \, .
\end{eqnarray}

\subsubsection{Back-reaction density}

We now reconsider the back-reaction density. At high masses, when the axion field is no longer coherent, the effective charge and current densities induced by the axion oscillate on scales much smaller than the gap size (and on times much smaller than the light crossing time of the gap). This implies that the average axion-induced charge and current densities tend toward zero in the limit that $m_a \rightarrow \infty$; an easy way to understand that this is the case is to recognize that the small scale oscillations in the electric field will leave the net voltage drop across the gap unchanged, and as such the axion is not expected to produce sizable changes to the gap collapse process. For any finite axion mass, however, there will exist a residual phase in the axion field that will not lead to a perfect cancellation. In a one-dimensional treatment, the spatio-temporal average (over the open phase of the gap) of $\nabla a$ yields
\vspace{0.95cm}
\begin{eqnarray}\label{Eq:gradLow}
\frac{1}{h_{\rm gap}\, \tau_{\rm open}} \int dt \int_{R_{\rm NS}}^{R_{\rm NS}+ h_{\rm gap}} dz \nabla a \, \,  \simeq  \frac{\sqrt{2\epsilon_a} \cos(m_a \, \tau_{\rm open} + \delta)}{m_a^2 \, h_{\rm gap} \tau_{\rm open}}\nonumber. \\ \vspace{0.1mm}
\end{eqnarray}
Taking $\tau_{\rm open} \sim  h_{\rm gap}$, we find that this limit yields a suppression of $\nabla a$ scaling as  $(\lambda_a / h_{\rm gap})^2$; equivalently, this corresponds to a back-reaction density in the large mass limit of
\begin{equation}
\epsilon_{\rm br} \simeq \xi^2 \frac{2 \, \Omega^2 m_a^4 h_{\rm gap}^4 }{ \, g_{a\gamma\gamma}^2},
\end{equation}
which differs from the case of small axion masses by a factor $\sim m_a^2 \, k_a^2 \, h_{\rm gap}^4$. Expressing the back-reaction density in terms of the Crab parameters, one finds
\begin{eqnarray}\label{eq:rho_br_highm}
\epsilon_{\rm br} \simeq 3 \times 10^{27} \, \rm \frac{GeV}{cm^3} \,  \Big(\frac{\xi}{0.1}\Big)^2 \, \left(\frac{m_{-6}^4}{g_{-12}^2} \right) \, \tilde{\Omega}^{-2/7} \, \tilde{B}_0^{-16/7} \nonumber. \\ \vspace{0.05mm}
\end{eqnarray}
Note that the extrapolation of Eq.~\ref{eq:rho_br_highm} and Eq.~7 in the main text to the threshold mass, \ie where $2\pi/m_a = h_{\rm gap}$, leads to a discontinuity -- this is  a result of the fact that Eq.~\ref{eq:rho_br_highm} has been derived under the assumption that $\lambda_a \ll h_{\rm gap}$. In order to generate smooth curves across the axion parameter space (as in Fig.~1 of the main text), we simply derive the mass $\bar{m}$ for which these curves cross (corresponding to $\bar{m} \sim 1.74 / h_{\rm gap}$), and define a piece-wise function for smaller and larger masses using  Eq.~7 in the main text and Eq.~\ref{eq:rho_br_highm}, respectively.

As an aside, one may wonder why the back-reaction density has been derived by averaging the axion gradient only over an the spatial direction oriented along the magnetic axis (and over time), but does not include an average over the radial direction. For young pulsars, acceleration and pair production is nearly one-dimensional, and thus averaging only over the z direction reflects the threshold at which local pair production would be effected, while including the radial direction instead reflects the point at which pair production would globally be effected. Including the radial direction would dramatically increase the back-reaction density -- we thus choose to adopt the more conservative threshold (this is conservative in that the  maximal density, which enters \eg the computation of the radio flux, is lower), which is set by restricting to the domain where the local dynamics of pair production begins to become altered.

\subsubsection{Energy dissipation rate}

Finally, we turn our attention to the energy dissipation rate. The expression derived in Eq.~\ref{eq:e_diss} corresponds to the energy loss arising from particle acceleration. At large masses, however, the electric field oscillates on small scales, and is not expected to induce any net voltage drop across the gap\footnote{Technically, there will always exist a small voltage drop arising from the imprecise cancellation of the phase. Following the derivation of the back-reaction density at high masses (but in this case averaging also over the radial direction), we determine that the rate of energy loss in Eq.~\ref{eq:e_diss} should scale as $\propto m_a^{-8}$ when $\lambda_a \lesssim h_{\rm gap}$. Bearing in mind that the production rate scales as $m_a^{-4}$, this implies the saturation density implied via particle acceleration should scale from Eq.~9 of the main body as $m_a^{-4}$. As shown below, this quickly implies that energy dissipation via particle acceleration is subdominant. }; as a consequence, the induced electric field will induce no net work on the system\footnote{There is a caveat that arises when the induced electric field becomes comparable to the intrinsic un-screened value of the electric field. Here, rapid acceleration can cause these particles to radiate high energy emission, and this can serve as an additional form of energy loss.}. Nevertheless, if the axion mass is larger than the local plasma frequency, one can also directly produce on-shell photons. At small masses, this process is blocked by the presence of the plasma -- this occurs since a dense plasma is produced within the oscillation timescale of the axion. At larger masses, however, the axion oscillates rapidly, and can produce on-shell radiation during the vacuum phase. In order to compute the rate of energy loss via electromagnetic emission we revisit the computation of the axion-induced electromagnetic fields, but focusing on the far-field (rather than near-field) limit. This is done by expanding the electromagnetic fields as $\vec{B} = \vec{B}_0 + \delta \vec{B}$ and $\vec{E} = \vec{E}_0 + \delta \vec{E}$, and re-deriving the wave equations in the absence of classical source terms, but in the presence of an axion background. The resultant solutions for the perturbed electromagnetic fields are given by 
\begin{subequations}\label{Eq:EMProduction}
\begin{eqnarray}\label{Eq:EMeqBack}
      (\partial_t^2 - \nabla^2) \delta\vec{E}&=&   \nabla \rho_a -\partial_t \, \vec{j}_a, \\[1ex]
    (\partial_t^2 - \nabla^2) \delta\vec{B}&=& - \nabla \times \vec{j}_a \,,
\end{eqnarray}
\end{subequations}
where for simplicity we have neglected the plasma frequency (\textit{i.e} strictly speaking we assume here $m_a \gg \omega_p$). These equations can be solved using Greens functions, which in the far-field limit  $|\vec{x} - \vec{x}^\prime| / |\vec{x}| \ll 1$ give
\begin{eqnarray}
\delta\vec{E}(\vec{x})&=& \frac{e^{-i \omega \, t + i \omega \vec{x}}}{4\pi |\vec{x}|} \int d^3\textbf{x}'e^{-i \omega \, \vec{x}\cdot \vec{x}'/|\vec{x}|}(\nabla 
 \rho_{a,x}(\vec{x}')\\ + i \omega \, \vec{j}_{a,x}(\vec{x}')), \nonumber
      \\[10pt]
    \delta\vec{B}(\vec{x})&=& - \frac{e^{-i \omega \, t + i \omega \vec{x}}}{4\pi |\vec{x}|}\int d^3\textbf{x}'e^{-i \omega \, \vec{x}\cdot \vec{x}'/|\vec{x}|} \, \nabla \times \vec{j}_{a,x}(\vec{x}'), \nonumber
\end{eqnarray}
where we introduced the notation $\vec{j}_{a, x} = e^{-i\omega t} \vec{j}_a$ and $\rho_{a, x} = e^{-i\omega t} \rho_a$. 
Integrating over the source yields 
\begin{subequations}
\begin{eqnarray}
      \delta\vec{E}(t, \vec{x})& \sim & \frac{e^{-i \omega \, t + i \omega \vec{x}}}{4\pi |\vec{x}| \, \omega^3} (i \, \vec{k} \rho_{a,x}(\vec{k}) + i \omega \, \vec{j}_{a,x}(\vec{k})),\\[2ex]
    \delta\vec{B}(t, \vec{x})& \sim& - \frac{e^{-i \omega \, t + i \omega \vec{x}}}{4\pi |\vec{x}| \, \omega^3}\, i \, \vec{k} \times \vec{j}_{a,x}(\vec{k}).
\end{eqnarray}
\end{subequations}
\\
Notice that ultimately the factor $1/\omega^3$ comes from the fact that in the high-mass regime the axion cloud is a sum of \textit{incoherent} dipoles, each with a volume $\sim 1/\omega^3$ much smaller than the gap volume.

Keeping only the leading order term in the non-relativistic limit, we see that the radiated power is  approximately given by
\begin{eqnarray}
   \biggl \langle \frac{dE}{dt}\biggr \rangle_{\rm a \rightarrow \gamma} &\simeq& 4\pi |\vec{x}|^2 \times \left[\delta\vec{E} \times \delta\vec{B} \right] \, \tau_{\rm duty} \nonumber \\ 
   &\sim& \frac{1}{4\pi} \, \frac{g_{a\gamma\gamma}^2 \, a^2 \, B_0^2}{\omega^2} \, \tau_{\rm duty}, 
\end{eqnarray}
which is then plugged in the expression of the observable flux density to obtain Eq.~10 of the main body. As before, we have incorporated a duty cycle to include the fact the vacuum phase only comprises a fractional part of the discharge process. Equating this expression with the energy injection rate, we find a saturation density in this regime to be
\begin{equation}
    \epsilon_{\rm sat} \simeq 8\times 10^{33}  \,\frac{\rm GeV}{\rm cm^3} \, m_{-5}^3 \, \tilde{B}_0^{10/7} \, \tilde{\Omega}^{24/7} \, .
\end{equation}
Clearly, energy dissipation is heavily suppressed at large masses, and thus we do not expect electromagnetic emission to impede the growth of the cloud.

\subsubsection{Radio Beaming}

As mentioned in the main text, the radio emission generated during the open phase of the gap is expected to be highly beamed along the magnetic axis. While the effect is similar to what occurs for the coherent radio emission produced by the pulsar itself, the mechanism driving the beaming is distinct. Here, we provide additional details on the beaming mechanism, and justify the choice of a $10^\circ$ opening angle used in the estimation of the flux density.

The coherent radio emission produced by the pulsar undergoes beaming because it is emitted from a highly boosted plasma. The axion-induced radio emission, on the other hand, is generated in the rest frame of the pulsar, and with only a minor preference in the direction of emission oriented perpendicular to the field lines (this can be seen by noting that $\delta \vec{E} \propto \vec{j} \propto \vec{B}$, while $\delta \vec{B} \propto \vec{k} \times \vec{j} \propto \vec{k} \times \vec{B}$). So how does emission which is emitted perpendicular to the field lines become beamed along the axis? The answer is a combination of refraction and reflection off of the dense plasma. The return currents, running along the last open field lines, have an enormously high density, and thus provide a reflective boundary which serves to confine the radiation in the open field lines. This alone, however, is not enough to provide heavily beamed emission, as the radiation would merely undergo a slow diffusion along the open field lines (which would not result in significant beaming). The second effect arises from diffraction off of the newly formed plasma within the gap itself. Recall that the radiation is only produced during the vacuum phase. As the primary $e^\pm$ particles pair produce, they generate a dense plasma (with an effective plasma frequency much above the radiation frequency) that is driven outward along the magnetic axis. This dense plasma acts like a relativistic bulldozer, driving the newly produced radiation along the field lines. 

These two processes suggest that the axion-induced radiation should be roughly concentrated in an angular region of the sky consistent with the angular opening of the gap itself. Taking the Crab pulsar as an example, we estimate the half-angle opening of the gap as $\theta_{radio} \sim  {\rm arctan}(r_{\rm pc} / R_{\rm NS}) \sim 4^\circ$, although this can be much smaller.

\begin{figure*}
    \centering
    \includegraphics[width = 0.32\textwidth]{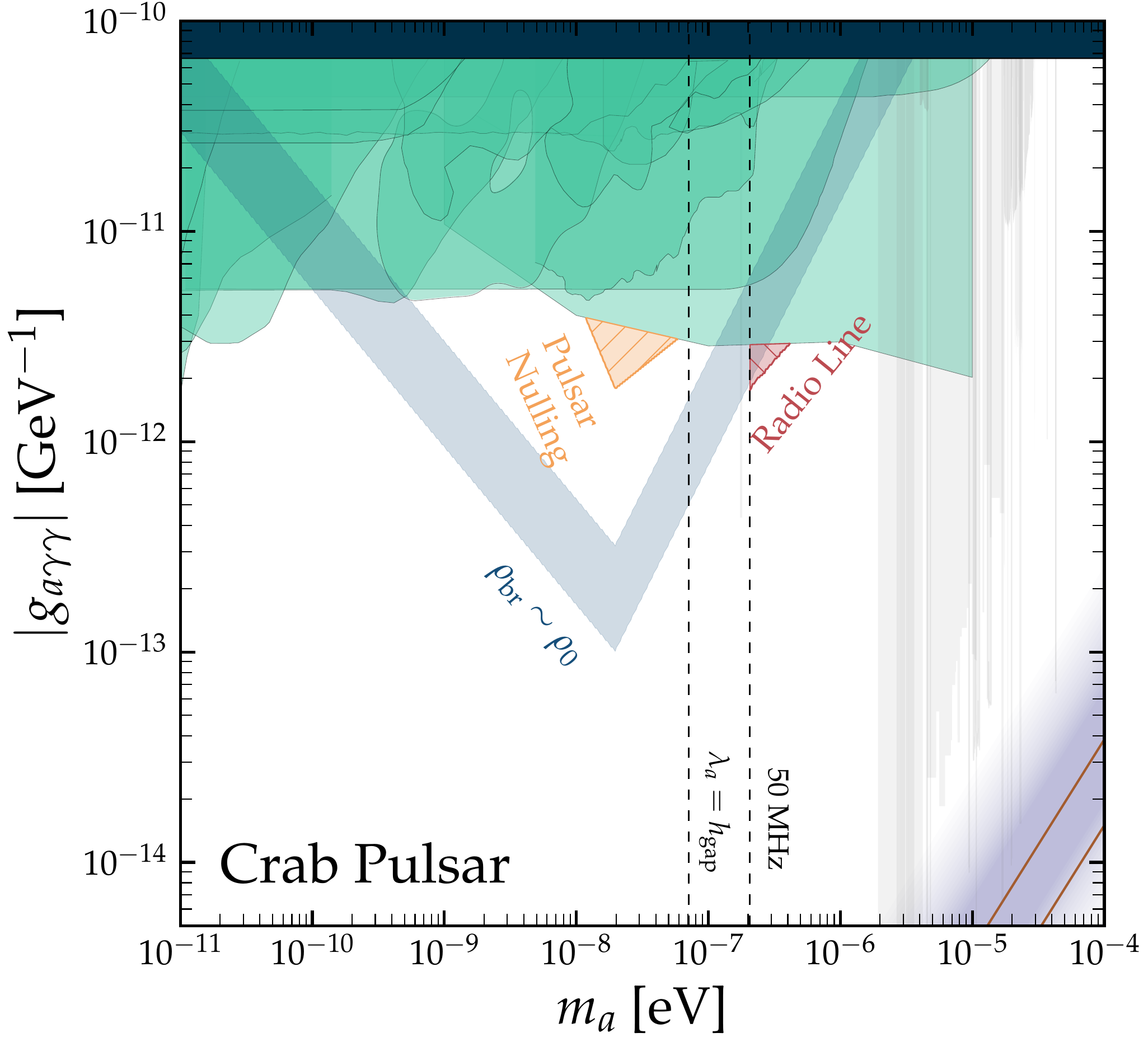}
    \includegraphics[width = 0.32\textwidth]{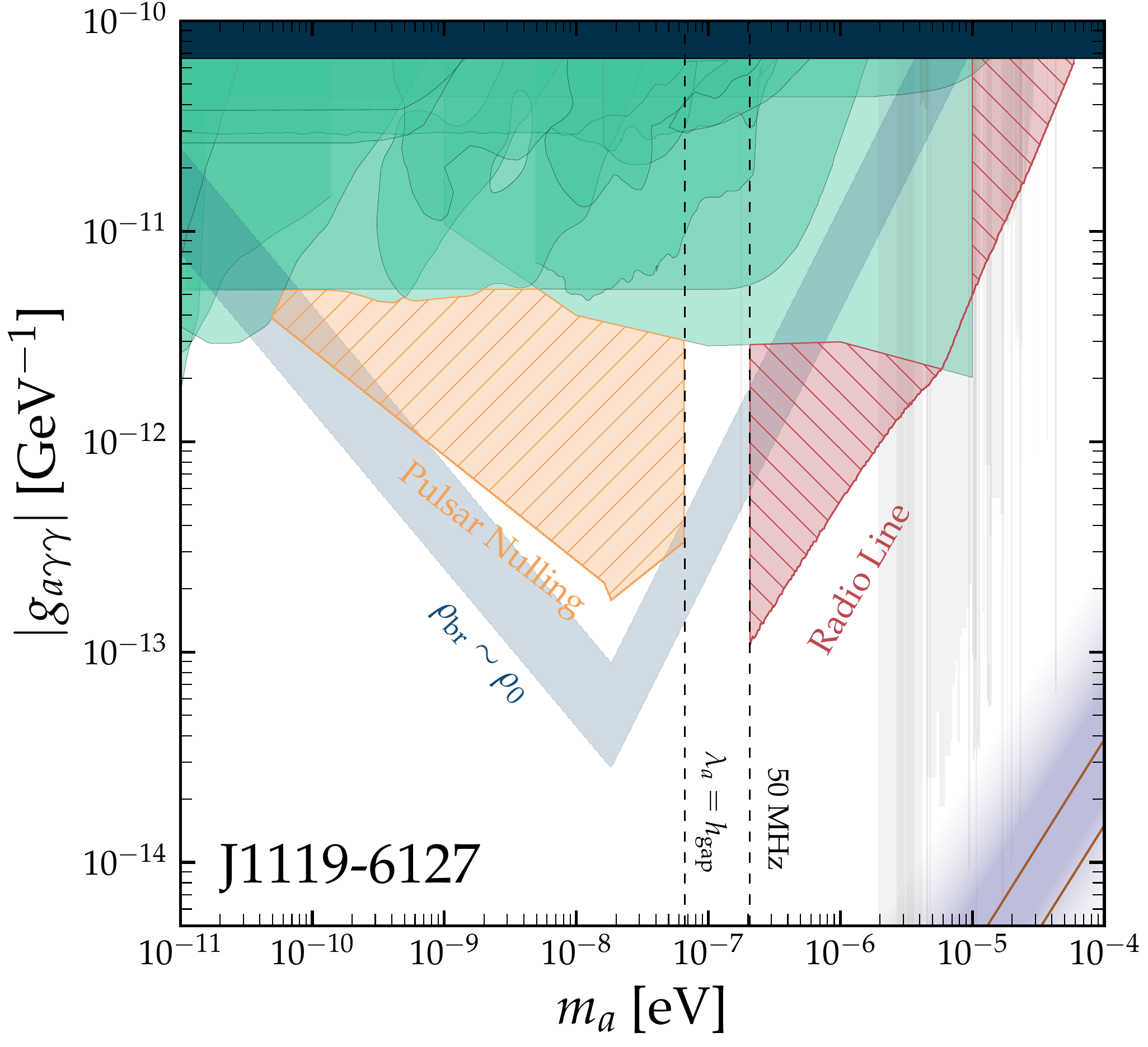}
    \includegraphics[width = 0.32\textwidth]{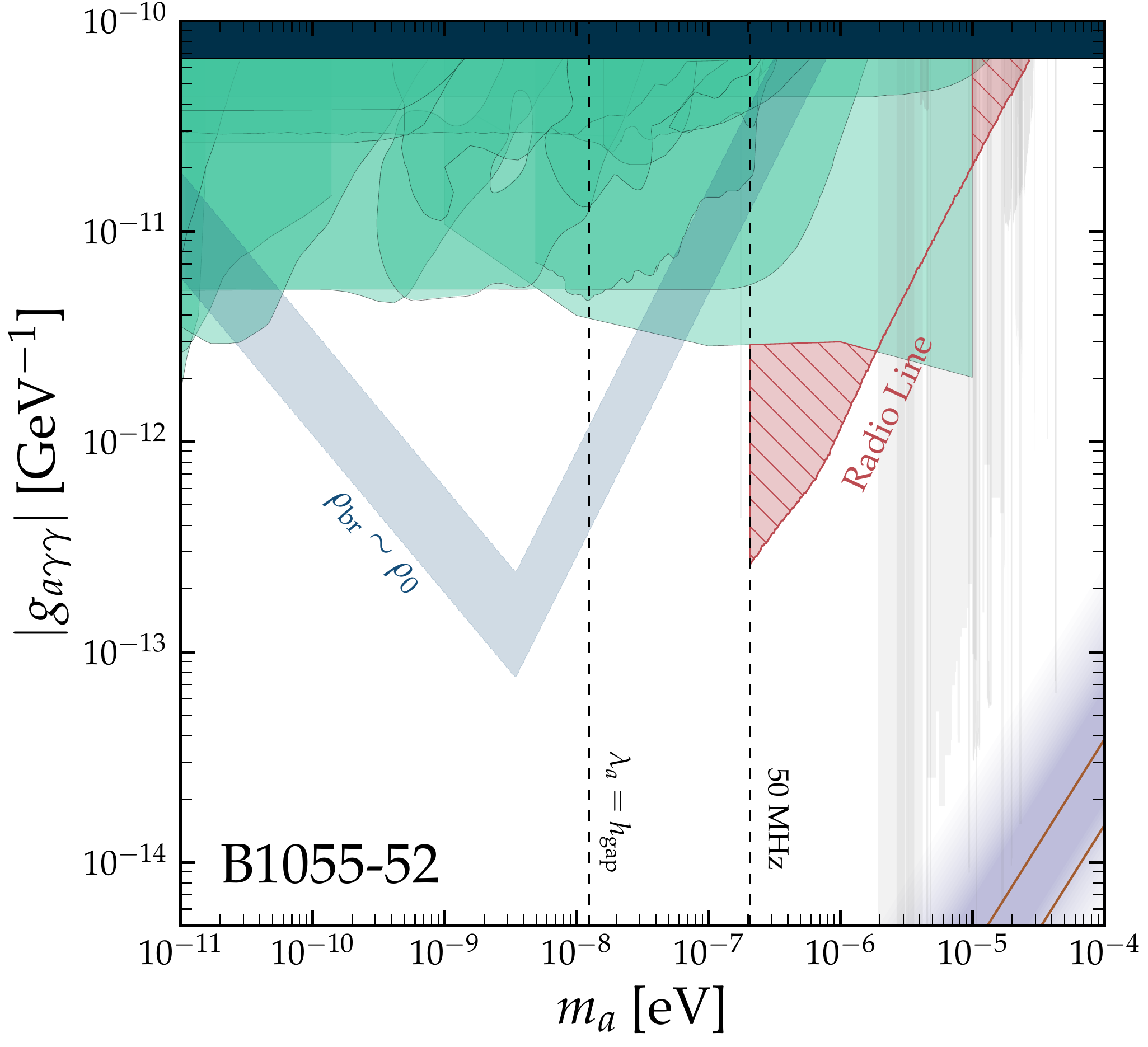}
    \caption{\label{fig:pulsars} Same as Fig.~1 of the main body, but showing the results for each of the three pulsars separately. Here, we also plot  the minimal coupling at which the projected axion density reaches $\epsilon_{\rm br}$ -- this is shown as a blue band, with the width of the band reflecting the impact of varying $\xi \in [0.1, 1]$. Vertical lines are drawn in each plot to indicate where $\lambda_a = h_{\rm gap}$ (above these curves we truncate the pulsar nulling contours) and the mass corresponding to $50$ MHz (below which we truncate the radio emission curves). Fig.~1 in the \textit{Letter} is the culmination of the bounds from all three of the shown pulsars.   }
\end{figure*}

\section{Magneto-rotational Spin-down}

We have assumed throughout this work that the pulsar properties today are reflective of those earlier in its lifetime -- this assumption is implicit in the linear scaling of the density profile with time. Neutron stars, however, are expected to undergo magneto-rotational spin-down, a process which can reduce the surface magnetic field strength and reduce the rotational frequency on timescales which are potentially shorter than the lifetime of the pulsars considered in this work. Here, we provide a very brief look at the impact of our current approximation.

The evolution of the pulsar period is determined by a combination of dipole radiation and plasma effects (see \eg \cite{Spitkovsky2006,Philippov2014}), and is described by the following equation:
\begin{eqnarray}
\dot{P} &=& \beta \frac{B^2}{P} \, (\kappa_0 + \kappa_1 \sin^2\chi) \, . \label{eqn:pevol} 
\end{eqnarray}
The scenario in which $\kappa_0 = 0, \kappa_1=2/3$ corresponds to the case of pure dipole radiation, while $\kappa_0 \sim \kappa_1 \sim 1$ has been inferred from numerical simulations of plasma-filled magnetospheres~\cite{Spitkovsky2006,Philippov2014}. Eq.~\ref{eqn:pevol} is often complemented by those governing the decay of the magnetic field and the evolution of the misalignment angle. The evolution of both of these quantities, however, is highly uncertain\footnote{ Refs.~\cite{Spitkovsky2006,Philippov2014} also investigate the evolution of the misalignment angle, which in general is expected to evolve toward zero. The evidence for this, however, is indirect, and is based on the observation that older pulsars tend to be more aligned. There are a few scenarios in which the evolution of the misalignment angle of a single pulsar has been inferred. One interesting example is the case of the Crab pulsar, which appears to be moving toward maximal misalignment (in disagreement with conventional reasoning), albeit slowly~\cite{lyne2013evolution,lyne201545}.}, and thus in what follows we choose to treat both of these quantities as constant over the course of the neutron star lifetime.  We expect this treatment to be conservative, as our choice amounts to an underestimation of the magnetic field and rotational frequency early in the life of the pulsar (the production rate being directly correlated with both of these quantities).

 By evolving Eq.~\ref{eqn:pevol} backward from the current pulsar parameters, one can look the evolution of the axion production over the history of the neutron star lifetime. For simplicity, we adopt the coefficients $\kappa_i$ corresponding to the vacuum scenario, and take the initial magnetic alignment angles for the Crab, B1055-52, and J119-6127 corresponding to $\chi \sim 40^\circ$, $\chi \sim 85^\circ$, and $\chi \sim 5^\circ$, respectively~\cite{lyne2013evolution,rookyard2015constraints}\footnote{There exists uncertainty in the true values of these parameters, but we simply adopt one representative value for illustrative purposes.}.

Having reconstructed $\Omega(t)$ for each pulsar, we follow the evolution of the density profile by replacing the linear time dependence of Eq.~6 in the main text with an integration. We find that in general, the axion  surface density  obtained using this procedure tends to be greater than the linear approximation given in Eq.~6 by a an $\mathcal{O}(1)$ factor, making the linear extrapolation conservative.

\section{Sensitivity Analysis}

Here, we outline the details of the sensitivity analysis leading to the projected contours shown in Fig.~1 of our \textit{Letter}. We focus in particular on three pulsars, including the Crab pulsar (chosen as the prototypical example of a young pulsar which has been studied for more than half a century), J1119-6127 (a pulsar near the Galactic Center that was chosen because it produces a larger ratio of $\epsilon_{\rm sat}/\epsilon_{\rm br}$), and a nearby pulsar B1055-52 (chosen so as to maximize Eq.~10 in the main text).  The details of each pulsar are included in Table~\ref{tab:pulsar}.

\begin{table*}[!ht]
    \centering
    \begin{tabular}{|l|l|l|l|l|l|l|l|l|l|l|l|}
    \hline
         Name / Property & Period [$s$] & $B_0$ [G] & $t_{\rm age}$ [yr] & d [kpc] & $S_{150}$  & $S_{600}$  & $S_{728}$  & $S_{800}$  & $S_{900}$  & $S_{1382}$  & $S_{3000}$  \\ \hline
        Crab  & 0.033 & 3.8e12 & 1e3 & 2 & 7500 [3800] & 211 [37] & - & - & 45 [10] & - & - \\ \hline
        J1119-6127 & 0.407 & 4.1e13 & 1.6e3 & 8.4 & - & - & 3 [0.4] & - & - & 1.09 [0.06] & 0.36 [0.03] \\ \hline
        B1055-52 & 0.197 & 1.09e12 & 5.3e5 & 0.093 & 310 [10] & - & 22 [5] & 20 [2] & - & 4.4 [0.6] & 1.4 [0.3] \\ \hline
    \end{tabular}
    \caption{\label{tab:pulsar}Properties of pulsars studied in this work. All flux densities are quoted in mJy, with $1\sigma$ uncertainties listed in brackets and the corresponding frequency in MHz given as lower labels (data is obtained from~\cite{jankowski2019utmost,bell2016time,jankowski2018spectral,weltevrede2011glitch}).}
\end{table*}

In order to determine the projected sensitivity to pulsar nulling, we  determine the value of the axion photon coupling for which $\epsilon_{\rm br}(\gagg) = \epsilon_{\rm sat}$ and $\epsilon_0(\gagg) \geq \epsilon_{\rm br}(\gagg)$, \ie we require the production of axions be sufficiently large to reach the back reaction density, and we require the back-reaction density be less than the saturation density. We only derive the back-reaction constraints for axion masses $m_a \leq 2\pi/h_{\rm gap}$, as it is unclear whether pulsar nulling will occur at higher masses. The derived limit is shown for each of the pulsars in the sample in Fig.~\ref{fig:pulsars}, and is derived assuming that $\xi = 0.1$ -- taking a larger value of $\xi$ increases $\epsilon_{\rm br} \propto \xi^2$, and thus can reduce the projected sensitivity. As expected, J119-6127 shows the strongest projected sensitivity to pulsar nulling. The broken power law feature seen in the pulsar nulling limit of J119-6127 is a consequence of the fact that the de Broglie wavelength of the axion becomes larger than the radial size of the gap, and thus the induced electric field (and consequently the energy dissipation rate) becomes increasingly suppressed.

In order to determine the projected sensitivity to the radio line we must first estimate the expected background from the pulsar's intrinsic flux.  Table~\ref{tab:pulsar} contains the observed flux density (along with corresponding $1\sigma$ uncertainties) for each of the pulsars in the sample at a variety of frequencies. Our procedure for determining the background-subtracted sensitivity for each pulsar is as follows. First, for each frequency bin we randomly draw $\mathcal{O}(10^3)$ samples about the mean observed value and taking the standard deviation to be the  $1\sigma$ uncertainty. For each set of samples (defined as containing a single random draw at each of the observed frequencies), we perform a power law fit to the flux density, \ie $S_\nu = A \times \, (\nu / 1.4 \, {\rm GHz})^\alpha$, and then evaluate the fit at the frequency of interest $\nu_a$. The collection of the samples $(S_{\nu, 1}(\nu_a), \cdots, S_{\nu, N}(\nu_a))$ represent estimates of the pulsar spectrum at that frequency. We assume the uncertainty is characterized by the standard deviation $\sigma_\nu$ across all of the samples, and define our observable flux density threshold to be $2\times \sigma_\nu$ (note that we set an absolute lower limit on the threshold using the minimal error recorded in Table~\ref{tab:pulsar}). We also impose an additional cut such that we only consider frequencies above 50 MHz, as lower frequencies can be efficiently absorbed before reaching Earth. The results are plotted for each pulsar in Fig.~\ref{fig:pulsars}. Recall that  B1055-22 has been selected to maximize the flux density in Eq.~10 of the \textit{Letter}; in Fig.~\ref{fig:pulsars} we see that the sensitivity is actually comparable to that of J119-6127 -- this arises because  Eq.~10 is only valid for axion-photon couplings above $g_{a\gamma\gamma}^{\rm min}$ (with the flux density scaling like $g_{a\gamma\gamma}^{4}$ for smaller couplings), and $g_{a\gamma\gamma}^{\rm min}$ for B1055-22 is larger than that of J1119-6127.

\section{An Aside on Resonant Transitions}

One may wonder whether the growth of the background plasma can trigger resonant axion-photon transitions. Here, we argue that the level crossing generated during the growth of the pair plasma is expected to proceed sufficiently fast that conventional calculations of the resonant conversion probability break down. In this regime the conversion is expected to be heavily suppressed, however without a more detailed formalism it is difficult to make detailed calculations which compare the energy losses with the non-resonant production of photons. 

In the non-relativistic limit, resonant axion-photon transitions occur when $m_a \simeq \omega_{p, {\rm eff}}$ where we have introduced the effective plasma frequency as the plasma frequency averaged over the inverse boost factor cubed, \ie $\omega_{p, {\rm eff}} \equiv \left< \omega_p / \gamma^{3/2} \right>$. The probability of axion photon transitions is roughly given by (neglecting order-one angular pre-factors) $p_{a\rightarrow\gamma} \sim g_{a\gamma\gamma}^2 \, B^2 \, (\partial_t \omega_{p, {\rm eff}})^{-1} / v$\footnote{Note that in standard scenarios the background plasma density is often treated as static, and thus the final term appears as a spatial (rather than temporal) derivative. Here, however, the background changes quickly, and thus the temporal derivative is expected to dominate. }. Note that the resonant timescale is roughly given by $t_{\rm res} \sim \sqrt{(\partial_t \omega_{p, {\rm eff}})^{-1}}$. The validity of the expressions outlined above are limited to the case where the background varies sufficiently slowly and level crossing is approximately linear in time. 

In the gap, the effective plasma frequency is expected to grow from an initially small value, $\omega_{p, {\rm eff}} \sim \omega_{GJ, 0} / (10^6)^3$, to the GJ value at the pole $\omega_{GJ, 0}$ (here, we have adopted a characteristic boost factor of $\gamma \sim 10^6$, with reasonable variations in this choice leading to negligible differences). The plasma is expected to grow exponentially, since plasma production proceeds iteratively over multiple generations with a growing multiplicity (and reduced boost factor) at each generation. Taking as an example an axion mass of $m_a \sim 10^{-6}$ eV, we find $t_{\rm res}$ to be much less than the oscillation timescale of the axion itself, illustrating that the conventional treatments of resonant mixing are invalid. 

We leave a more detailed understanding of whether resonant transitions could contribute to additional photon production to future work, and simply note here that if resonances are capable of enhancing photon production (beyond what is generated non-resonantly), the dominant effect will be an enhancement the radio flux, which would imply that the treatment outlined here is conservative. 


\bibliographystyle{bibi}
\bibliography{biblio}

\end{document}